\documentclass[useAMS,usenatbib]{mn2e}
\usepackage{graphics,graphicx}
\usepackage{amssymb}
\usepackage{latexsym}
\usepackage{color}
\usepackage{epstopdf}
\usepackage{marvosym}

\begin{document}

\title[Electron-impact excitation of diatomic hydride
  cations]{Electron-impact excitation of diatomic hydride cations \\
  I: HeH$^+$, CH$^+$, ArH$^+$} \author[James R Hamilton, Alexandre Faure
  and Jonathan Tennyson]{James R Hamilton$^1$\thanks{E-mail:
    james.hamilton@ucl.ac.uk}, Alexandre Faure$^2$\thanks{E-mail:
    alexandre.faure@obs.ujf-grenoble.fr} and Jonathan
  Tennyson$^1$\thanks{E-mail: j.tennyson@ucl.ac.uk}\\ $^1$ Department
  of Physics and Astronomy, University College, London, Gower St.,
  London WC1E 6BT, UK\\ $^2$ UJF-Grenoble 1 / CNRS-INSU, Institut de
  Plan\'etologie et d'Astrophysique de Grenoble (IPAG) UMR 5274,
  Grenoble, F-38041, France}

\date{Accepted ? Received ?}

\maketitle

\begin{abstract}

{\bf R}-matrix calculations combined with the adiabatic nuclei
approximation are used to compute electron-impact rotiational
excitation rates for three closed-shell diatomic cations, HeH$^+$,
CH$^+$, ArH$^+$. Comparisons with previous studies show that an
improved treatment of threshold effects leads to significant changes
in the low temperature rates; furthermore the new calculations suggest
that excitation of CH$^+$ is dominated by $\Delta J =1$ transitions as
is expected for cations with a large dipole moment. A model for
ArH$^+$ excitation in the Crab Nebula is presented which gives results
consistent with the observations for electron densities in the range
$2-3\times 10^3$~cm$^{-3}$.

\end{abstract}
\textit{molecular processes; astronomical data bases: planetary nebulae: general; ISM: molecules}

\section{Introduction}

Because hydrogen is much more abundant than any other element in the
interstellar medium (ISM), hydrides are the first molecules to
form. They therefore constitute significant reservoirs of heavy
elements, as well as sensitive tests of interstellar chemistry
networks and of the ambient physical conditions. 
Hydrides are also employed to trace specific aspects of
interstellar environments. For example, in diffuse interstellar
clouds, hydroxyl and water cations, OH$^+$ and H$_2$O$^+$, are useful
probes of the cosmic ray ionization rate \citep{Indriolo2012}, which
controls the abundance of several interstellar molecules. Hydrogen
fluoride, HF, is formed by the direct reaction of H$_2$ with F and can
be employed as a proxy for molecular hydrogen, see
\citet{Indriolo2013} and references therein. The methylidyne and
mercapto cations, CH$^+$ and SH$^+$, are good tracers of warm regions,
heated by shocks or the dissipation of turbulence, where endothermic
reactions can occur, e.g.  \citet{Godard2012}. The argonium cation,
ArH$^+$, was the first noble gas molecular ion detected
in space \citep{Barlow2013} and it has been shown to be a tracer of almost purely atomic gas
in the diffuse ISM \citep{Schilke2014}. 
As discussed by \citet{Jimenez-Redondo2014},
molecular ions, such as
ArH$^+$, are also important constituents of cold terrestrial plasmas.

In diffuse clouds, hydrides can be observed through
optical/ultraviolet absorption lines superposed on the spectra of
background stars. At longer wavelengths, rotational transitions of
hydrides can be observed in both absorption and emission, providing a
large sampling of molecular regions, e.g. \citet{Gerin2010}. In
contrast to absorption line studies, which provide simple and reliable
estimates of column densities, the interpretation of emission lines is
much more problematic. The density of the dominant colliders (H, He,
H$_2$ and electrons) is indeed often close to the critical
density\footnote{The critical density defines, for a given transition
  or level, the density at which the spontaneous radiative rate equals
  the collisional rate.}. In such conditions, deviations from
local-thermodynamic equilibrium (LTE) are expected and detailed
radiative transfer modeling is necessary to derive both the physical
conditions (density and temperature) and the column
densities. Radiative transfer calculations in turn require a good
knowledge of rate coefficients for collisional excitation. Despite
recent experimental progress, these coefficients are difficult to
measure and theory is needed to provide comprehensive sets of
collisional data.

Significant progress was made in recent years in computing collisional
rates for interstellar molecules, see \citet{Roueff2013} for a review
on collisions with H, He and H$_2$. In regions where the electron
fraction, $n_e/n_{\rm H}$, is larger than $\sim 10^{-5}-10^{-4}$,
electron-impact excitation can compete or even dominate over the
neutrals (H, He and H$_2$). For example, in the photon-dominated
region (PDR) associated with the Orion bar, \citet{Tak2012} have shown
that the excitation of HF is driven by electron collisions with
densities $n_e\sim 10$~cm$^{-3}$. A similar result was obtained
recently for OH$^+$ emission \citep{Tak2013,Gomez-Carrasco2014}.
Electron-impact excitation of diatomic hydride cations thus appears as
a crucial process in the ionized regions of the molecular ISM. 

  To date, five diatomic hydride cations
have been detected in the ISM: CH$^+$ \citep{Douglas1941}, OH$^+$
\citep{Wyrowski2010}, SH$^+$ \citep{Menten2011}, HCl$^+$
\citep{DeLuca2012} and very recently ArH$^+$ \citep{Barlow2013}. Thus
far searches for HeH$^+$ have not been successful
\citep{Moorhead1988,jt206,Zinchenko2011}, except for a tentative
detection in the remnant of supernova 1987A \citep{jt110}. We note
that HeH$^+$ is thought to be the first molecular species to appear in
the Universe \citep{Lepp2002}.

At interstellar medium temperatures, the process of electron
impact rotational excitation of hydrides should have
a clearly observable effect. However,
the process of dissociative recombnation competes with excitation;
results presented later in this work show that the cross
section for the rotational excitation of CH$^+$ in the dominant
channel $J = 0 $ to $J = 1$ is 10 to 100 times higher than the
summed experimental dissociative recombination cross section of CH$^+$
at the same temperature \citep{Amitay1996}. The cross section for
rotational excitation for the same channel of HeH$^+$ is also found to
be similarly greater than the
summed theoretical dissociative recombination cross sections computed by
\citet{Sarpal1994}, albeit not at resonance regions in the dissiciative
recombination cross section which nonetheless never exceeded the
magnitude of the rotational excitation cross section.

In this work, electron-impact rotational-excitation
 rate coefficients are calculated for
HeH$^+$, CH$^+$ and ArH$^+$, which all have a ground electronic state
of symmetry $^1\Sigma^+$. The {\bf R}-matrix method is combined with
the adiabatic-nuclei-rotation (ANR) approximation to obtain rotational
cross sections at electron energies below 5~eV. It should be noted
that previous {\bf R}-matrix studies on HeH$^+$ \citep{jt222} and
CH$^+$ \citep{jt237} were carried out but, particularly for the CH$^+$
problem, the accuracy was compromised by the use of reduced treatments
of polarisation and a small close-coupling expansions. In both cases
the use of an improper threshold correction made the results less
accurate at low electron temperatures. These limitations are overcome
in the present work, as detailed below. In Section~2, the {\bf
  R}-matrix calculations are described and the procedure used to
derive the cross-sections is briefly introduced. In Section 3, we
present and discuss the calculated rate coefficients. A model for the
excitation of ArH$^+$ in the Crab Nebula is also presented in
Section~4. Conclusions are summarized in Section~5.

\section{Calculations}

\subsection{R-matrix calculations}

In this paper we present results for the electron collision
calculations with molecular ions ArH$^+$, HeH$^+$ and CH$^+$. These
calculations employ the polyatomic {\bf R}-matrix method and are adapted to
calculate the rotational excitation cross sections for specific
transitions of the angular momentum quantum number, $J$, of the
molecules. The {\bf R}-matrix method is a sophisticated way of
performing electron scattering calculations. For a full derivation of
the {\bf R}-matrix method and its application in electron scattering
see \citet{jt474}. Here we used the  Quantemol-N expert system \citep{jt416}
to run the UK molecular {\bf R}-matrix codes (UKRMol) \citep{jt518}.

The calculations for all three molecules used {\bf R}-matrix radii of
13~a$_o$ and Gaussian Type Orbitals (GTOs) to represent both the
target and the continuum. Continuum orbitals up to g-wave ($\ell \leq
4$) were explicitly included in the calculations.  The initial
calculations for molecules were carried out using C$_{2v}$
symmetry. In order to calculate the rotational excitation cross
sections of the molecules, however, the {\bf T}-matrices were
tranformed into the C$_{\infty v}$ point group, the natural point
group of the diatomic molecules. The transformed {\bf T}-matrices
formed the input to electron-impact rotational excitation code
\texttt{ROTIONS} \citep{jt226}.

The technical details of the calculations are described below.

\subsubsection{ArH$^+$}

The ArH$^+$ target was represented using an augmented aug-cc-pVTZ GTO
basis set. The use of augmented basis sets improves the treatment of 
the more diffuse orbitals of the excited states in the calculation.  
The quantum chemistry program
\texttt{MOLPRO} \citep{molpro} was used to do initial Hartree-Fock calculations
on the ArH$^+$ ground state using an aug-cc-pVTZ basis set. These
ground state orbitals were then used in the {\bf R}-matrix
calculation.  The target was represented using complete active space
configuration interaction (CAS-CI) treatment involving freezing 10
electrons in 1$\sigma^2$ 2$\sigma^2$ 3$\sigma^2$ 1$\pi^4$ orbitals and
distributing the remaining 8 electrons in the orbitals: [4$\sigma$
  5$\sigma$ 2$\pi$ 6$\sigma$ 7$\sigma$ 8$\sigma$ 3$\pi$
  9$\sigma$]$^{8}$. Five states per symmetry generated from the
initial CAS-CI calculation were used in the final close-coupling
calculation with an excitation energy cut off of 30~eV. This helps to
converge the polarisation effects introduced by the coupling to
excited states of the molecule.

In the Earth's atmosphere the isotope of argon in overwhelming
abundance is \textsuperscript{40}Ar (exact mass 39.96238 u) as this
isotope is created by the decay of \textsuperscript{40}K. However
astronomically the major isotope of argon is \textsuperscript{36}Ar
(exact mass 35.96755 u). A third stable isotope of argon is know to exist in nature, 
\textsuperscript{38}Ar (exact mass 37.9627 u). \textsuperscript{38}Ar is the 
least abundant natural isotope and is not significant terrestrially or astronomically. 
The dipole moment is important when
calculating rotational excitation cross sections and, for ionic
molecules, the dipole moment depends upon the molecular
centre-of-mass. This means that the dipole differs between
isotopologues.

Our calculated dipole moment for \textsuperscript{36}ArH$^+$ is $\mu$
= 2.1425 D, for \textsuperscript{38}ArH$^+$ $\mu$
= 2.1517 D and for \textsuperscript{40}ArH$^+$ $\mu$ = 2.1595 D, with
rotational constant $B= 10.27$~cm\textsuperscript{-1} for all three. The
best available theoretical value for \textsuperscript{40}ArH$^+$ at its equilibrium geometry is
$\mu$ = 2.177 D \citep{Cheng2007}. By applying the same ratio to this
latter value as was found with our calculations, our best estimates are
$\mu$ = 2.169 D for \textsuperscript{38}ArH$^+$ and $\mu$ = 2.160 D for \textsuperscript{36}ArH$^+$. The
experimental value for the rotational constant of
\textsuperscript{36}ArH$^+$, from the Cologne Database for Molecular
Spectroscopy (CDMS), is $B$ = 10.30 cm\textsuperscript{-1}
\citep{Muller2005} and this is taken to be the value for \textsuperscript{38}ArH$^+$ and
\textsuperscript{40}ArH$^+$ also.
We note that
the rotational constants for the three isotopologues actually
differ by $~ 0.01 cm^{-1}$ but this effect, as well as corrections due to
centrifugal distorsion,
can be ignored in our treatment as they only influence the
threshold correction we employ (see below) and this
assumption itself introduces much
larger uncertainties at low temperature. On the other hand,
the small change in dipole values was taken into accound
since the Coulomb-Born correction scales with the square of
the dipole.

\subsubsection{HeH$^+$}

The HeH$^+$ target was represented using a cc-pVTZ GTO basis set.
Augmented basis sets gave result which showed artificial structure as
function of electron collision energy. This is a signature of problems
with linear dependence and probably caused by an excessive overlap
between the augmented functions and the molecule continuum functions.
The ground state electron occupation of HeH$^+$ is 1$\sigma^2$.  As
HeH$^+$ is a two electron system both its target wavefunction and the
scattering wavefunction were represented with a full CI treatment,
{\it i.e.} by allowing the electrons to freely occupy all orbitals.
All generated states up to 42~eV were retained in the final
close-coupling calculation.

Our calculated HeH\textsuperscript{+} equilibrium geometry dipole moment and rotational constant
are $\mu$ = 1.7168 D and $B$ = 35.241 cm\textsuperscript{-1}. The best
available theoretical value for the dipole is $\mu$ = 1.664 D
\citep{Pavanello2005} while the CDMS rotational constant is $B$ =
33.559 cm\textsuperscript{-1} \citep{Muller2005}.

\subsubsection{CH$^+$}

The previous study on CH$^+$ by \citet{jt237} was performed using
scattering wavefunctions computed using the older linear molecular
R-matrix code \citep{jt225} which employs Slater Type Orbitals, and a
rather crude representation of the lowest 6 target states in a
close-coupling expansion. Numerical and computational meant that only
rather simple calculations could be performed which failed to give a
full treatment of polarisation effects.  As demonstrated by a recent
study of bound and continuum states of N$_2$ \citep{jt560,jt574}, the
UKRMol code and GTO basis functions employed here are capable of
giving a fully converged treament of the problem. Furthermore modern
computers make the use of extensive close-coupling expansions
straightforward; such expansions are important for reprsenting
polarisation effects \citep{jt468}.

The CH$^+$ target was represented using a cc-pVTZ GTO basis
set. Again, augmented basis sets did not give smooth results.  The
ground state of CH$^+$ has the configuration [1$\sigma$ 2$\sigma$
  3$\sigma$]$^6$. The target was represented using CAS-CI treatment
freezing the lowest energy 2$\sigma^2$ orbital and placing the highest
4 electrons in orbitals [2$\sigma$ 3$\sigma$ 1$\pi$ 4$\sigma$
  5$\sigma$ 2$\pi$]$^4$. 51 excited states used with an
excitation energy cut off of 25~eV.

Two isotopes of CH\textsuperscript{+} are considered in this paper: 
\textsuperscript{12}CH$^+$ and \textsuperscript{13}CH$^+$ the calculated
equilibrium geometry dipole moment and rotational constant of \textsuperscript{12}CH$^+$ and \textsuperscript{13}CH$^+$
are the same to five significant figures and are $\mu$ = 1.6711 D and $B$ = 14.454 cm\textsuperscript{-1} respecively. The best
available experimental value for the \textsuperscript{12}CH$^+$ dipole is $\mu$ = 1.683 D
\citep{Cheng2007} which was used to calculated rotational excitation cross sections for 
\textsuperscript{12}CH$^+$ and \textsuperscript{13}CH$^+$. The CDMS rotational constant for 
$^{12}$CH$^+$ $B$ = 13.931 cm\textsuperscript{-1}\citep{Muller2005} 
was used for both $^{12}$CH$^+$ and $^{13}$CH$^+$.

\subsubsection{Cross-sections and rate coefficients}

Each of the above calculations produces four fixed-nuclei {\bf
  T}-matrices for each molecule (one for each irreducible representation
in C$_{\rm 2v}$ symmetry) for electron collision energy considered.
 These {\bf T}-matrices are used to
calculate the rotational excitation cross sections for the molecules
using program \texttt{ROTIONS} \citep{jt226}. \texttt{ROTIONS}
computes the rotational excitation cross sections for each transition
of the rotational angular momentum quantum number $J$. It is also
useful to look at the cross sections in terms of $\Delta J$, the
change in the angular momentum quantum number.

$\Delta J = 1$ transitions are strongly influenced by the long-range
effect of the dipole moment and \texttt{ROTIONS} employs the
Coulomb-Born approximation to include the contributions of higher
partial waves ($\ell > 4$) to the cross sections. These long-range
effects have been demonstrated to be unimportant for other transitions
\citep{jt271}.  The best available value of the dipole moment was used
in these calculations.

The adiabatic nuclear rotation (ANR) method was employed to calculate
the cross sections. The ANR approximation is valid when the electron
collision time is less than the period of rotational motion. The
suitability of the ANR method to diatomic, molecular ions is explained
in \cite{Chang1970}. As explained by \citet{jt396} the ANR method becomes 
invalid close to a rotational threshold because it neglects the 
rotational Hamiltonian. A kinetic scaling method, scaling with a kinematic 
momentum ratio solves this problem for neutral molecules but not, however, for 
ions, see \citet{jt396} and references therein. This kinetic-scaling method has been 
employed in the previous {\bf R}-matrix studies on HeH$^+$ and CH$^+$ and the 
cross sections published in these studies can be said, therefore, to be erroneous 
in the vicinity of the rotational threshold.
For this publication the threshold energy for rotational excitation
cross sections was calculated from the experimental rotational
constant of the molecule and cross sections below this threshold were
simply set to zero. This form of threshold ``correction'' is in accord with the
Wigner law and has been
shown to be much more accurate \citep{jt396} than the more complicated
kinetic-scaling employed in the previous {\bf R}-matrix studies on
HeH$^+$ and CH$^+$. Note that closed-channel effects are neglected but
these are expected to be small for polar ions \citep{jt396}. This
methodology was shown to work well in the only available experimental
test of electron impact rotational (de-)excitation rates \citep{jt465}.

In practice, we have considered rotational transitions between levels
with $J\leq 10$. Transitions were however restricted to $\Delta J\leq
8$ owing to the limited number of partial waves in the {\bf
  T}-matrices ($\ell \leq 4$). Excitation cross sections were computed
for collision energies $E_{coll}$ in the range 0.01-5~eV. For
transitions with a rotational threshold below 0.01 eV, cross sections
were extrapolated down to the threshold using a $1/E_{coll}$
(Wigner's) law, as recommended by \cite{jt396}. Assuming that the
electron velocity distribution is Maxwellian, rate coefficients for
excitation transitions were obtained for temperatures in the range
1  -- 3000~K. Going to higher temperatures would require the 
inclusion of $\Delta J\geq8$ transitions for the results to converge. 
Furthermore the energy range 1  -- 3000~K 
is considered complete for the interstellar medium. 
De-excitation rate coefficients were obtained using the
detailed balance relation.

\section{Results}
The supplementary data associated with this paper includes rotational excitation
and de-excitation rate coefficients for HeH$^+$, the two isotopes of CH$^+$: 
$^{12}$CH$^+$ and $^{13}$CH$^+$,  and the three isotopes of ArH$^+$: $^{36}$ArH$^+$, $^{38}$ArH$^+$ and $^{40}$ArH$^+$.
The datasets are published for transition with starting values of J = 0 - 11 which is 
complete for the temperature range 0 - 3000 K with $\Delta$J = 1 - 8.

We start by considering results for electron-impact rotational
excitation of HeH$^+$ and $^{12}$CH$^+$ (henceforth in this paper 
to be referred to simply as CH$^+$) since there are previous studies on
these systems against which we can compare.

\subsection{HeH$^+$}  

\begin{figure}
\begin{center}
	\includegraphics*[scale=0.35]{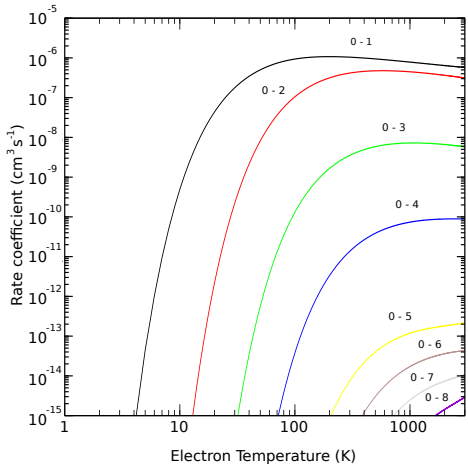}
\end{center}
	\caption{Rate coefficients for rotational excitation of
          HeH$^+$ from the ground state ($J$=0) to the lowest eight
          excited states.}
	\label{fig:hehp}
\end{figure}
\begin{figure}
\begin{center}
	\includegraphics*[scale=0.35]{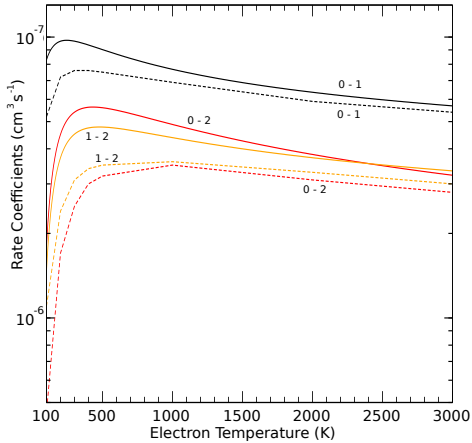}
\end{center}
	\caption{Rate coefficients for rotational excitation of HeH$^+$
          from \citet{jt222} (\Kutline) against our results
          (\textemdash).}
	\label{fig:hehp-1998}
\end{figure}

Figure~\ref{fig:hehp} shows our rate coefficients for rotational
excitation of HeH$^+$. As expected the $\Delta J = 1$ excitation is
the dominant process, particularly at low temperatures. However, as
observed by \citet{jt222}, the $\Delta J = 2$ excitation rate
approaches that of $\Delta J = 1$ for electron temperatures above
100~K, where it differs by only about a factor of two. This should
make the HeH$^+$ $J = 2 - 1$ emission line at 74.7848 $\mu$m
\citep{Matsushima1997} of similar strength to the $J = 1 - 0$ line at 149.137
$\mu$m.  Although both these far-infrared lines lie at wavelengths
which are difficult to study from the ground, the possibility of
observing two lines should significantly enhance the prospect of
observing this elusive but fundamental species.

Figure~\ref{fig:hehp-1998} compares the rate coefficients for key
electron-impact HeH$^+$ excitation transitions with the previous
results of \citet{jt222}. The agreement is generally very good. This
is not surprising since the electronic simplicity of the HeH$^+$
system meant that the earlier study was already based on excellent
electronic wavefunctions \citep{jt103} which had already been
demonstrated by detailed comparisons with experiment for the HeH
Rydberg molecule \citep{jt106}. The largest differences between the
previous study and our new results occur at low temperatures where our
rate coefficients are increased. This is due to our improved treatment
of threshold effects in the ANR approximation. For this reason we
believe our new results to be more accurate.

\subsection{CH$^+$}

\begin{figure}
\begin{center}
	\includegraphics*[scale=0.35]{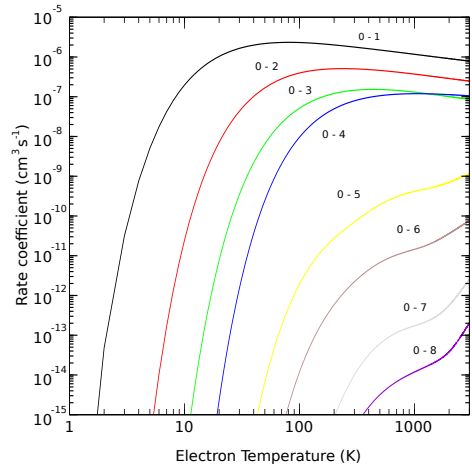}
\end{center}
	\caption{Rate coefficients for rotational excitation of CH$^+$
          from the ground ($J=0$) state to the lowest eight excited
          states.}
	\label{fig:chp}
\end{figure}

\begin{figure}
\begin{center}
	\includegraphics*[scale=0.35]{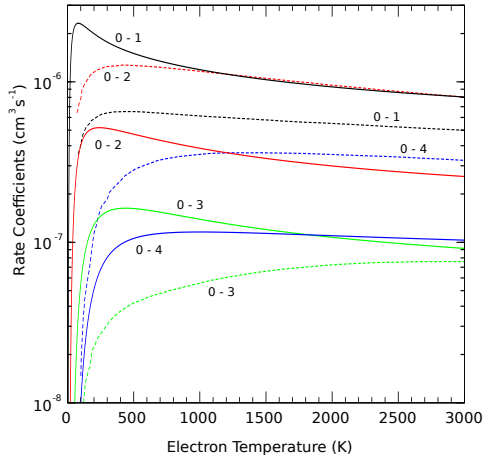}
\end{center}
	\caption{Rate coefficients for rotational excitation of CH$^+$
          from \citet{jt237} (\Kutline) against our results
          (\textemdash).}
	\label{fig:chp-1999-2014}
\end{figure}

The situation for CH$^+$ is somewhat different in that our new results
differ significantly from the previous
predictions. Figure~\ref{fig:chp} summarises our results. As might be
expected, given that CH$^+$ has a large permanent dipole, we find that
the excitation process is dominated by $\Delta J = 1$ transitions.

These rate coefficient can be compared with those calculated in a
previous {\bf R}-matrix study by \citet{jt237}. These authors found
the rotational excitation cross sections for the $\Delta J = 2$
transitions to be greater than $\Delta J = 1$ transitions. A
comparison with these results is shown in
Figure~\ref{fig:chp-1999-2014}. CH$^+$ is a difficult system to
construct accurate electron collision wavefunctions for, due to
presence of low-lying electronically excited states
\citep{jt63,jt499}. We suggest that the previous study overestimates
the role of $\Delta J = 2$ transitions for this strongly dipolar
system and our new results should be adopted.

Our rotational excitation cross sections for $\Delta J = 3$ and
$\Delta J = 4$ transitions in CH$^+$ agree quite well (within a factor
of 2) with those of \citet{jt237} at high temperatures where the
$\Delta J = 4$ cross section is greater. The agreement is however less
good at low temperatures where we find the $\Delta J = 3$ transition
to be more important; this change is largely due to our improved
treatment of threshold effects. Our new calculations also shows the
$\Delta J = 6$ transition to be much lower than previously shown; this
is again as expected.

\subsection{ArH$^+$}

\begin{figure}
\begin{center}
	\includegraphics*[scale=0.35]{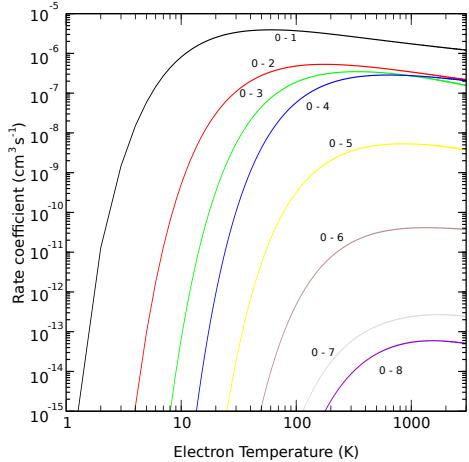}
\end{center}
	\caption{Rate coefficients for rotational excitation of
          $^{36}$ArH$^+$ from the ground state ($J=0$) to the lowest
          eight excited states}
	\label{fig:arhp36}
\end{figure}

\begin{figure}
\begin{center}
	\includegraphics*[scale=0.35]{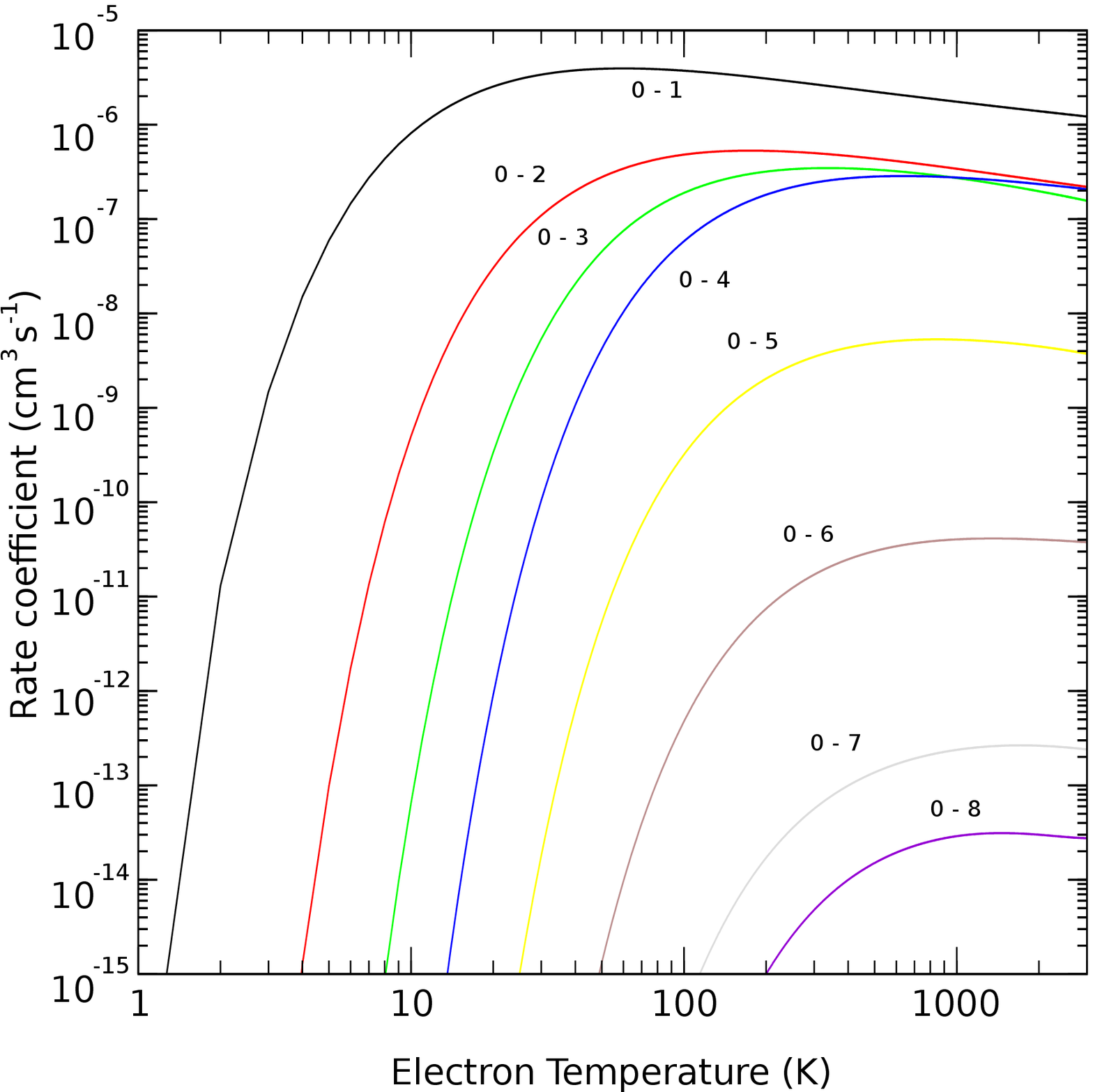}
\end{center}
	\caption{Rate coefficients for rotational excitation of
          $^{38}$ArH$^+$ from the ground state ($J=0$) to the lowest
          eight excited states}
	\label{fig:arhp38}
\end{figure}

\begin{figure}
\begin{center}
	\includegraphics*[scale=0.35]{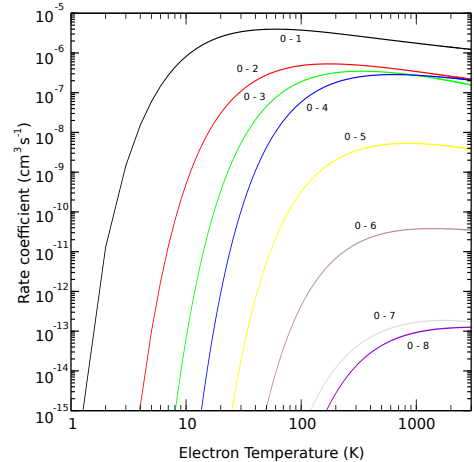}
\end{center}
	\caption{Rate coefficients for rotational excitation of
          $^{40}$ArH$^+$ from the ground state ($J=0$) to the lowest
          eight excited states}
	\label{fig:arhp40}
\end{figure}

Figure~\ref{fig:arhp36} presents rate coefficients for electron-impact
rotational excitation of \textsuperscript{36}ArH$^+$ from its rotational ground state.
Figures~\ref{fig:arhp38} and \ref{fig:arhp40} present rate coefficients for \textsuperscript{38}ArH$^+$
and \textsuperscript{40}ArH$^+$ respectively. As
expected the processes are dominated by $\Delta J = 1$
transitions. Interestingly, above 100~K the rates for for $\Delta J =
2$, $\Delta J = 3$ and $\Delta J = 4$ transitions become increasingly
similar. Rates for higher excitations are all negligible.

\section{{\protect{ArH$^+$}} excitation in the Crab Nebula}

The first detection of $^{36}$ArH$^+$ in space was reported by
\cite{Barlow2013} using the Herschel Space Observatory. The $j=1-0$
and $2-1$ rotational lines were identified in emission at several
positions in the Crab Nebula. This supernova remnant is known to
contain filaments where enhanced emission of both H$_2$ and Ar$^+$ is
observed. From the modeling of the two rotational lines,
\cite{Barlow2013} concluded that the likely excitation mechanism is
electron collisions in regions where the electron density $n_e$ is a
few$\times 100 $~cm$^{-3}$ and the temperature is about
3000~K. However, as collisional data were not available for ArH$^+$,
\cite{Barlow2013} employed the CH$^+$+e$^-$ collisional rates of
\cite{jt237} in place of those for ArH$^+$+e$^-$. The $^{36}$ArH$^+$
column density was estimated as 10$^{12}-10^{13}$~cm$^{-2}$, with
lower limits for the isotopic ratios $^{36}$ArH$^+$/$^{38}$ArH$^+>2$
and $^{36}$ArH$^+$/$^{40}$ArH$^+>4$. As can be seen from
  Figs. 4 and 5, the CH$^+$+e$^-$ collisional rates of \cite{jt237}
  differ significantly from those computed by us for ArH$^+$+e$^-$, so
  that different excitation conditions and/or column densities are
  expected.

A new model for the excitation of ArH$^+$ in the Crab Nebula is
presented here using the above {\bf R}-matrix electron-impact
excitation rates and the radiative rates taken from the CDMS
\citep{Muller2005}. Radiative transfer calculations were performed
with the \texttt{RADEX} code \citep{Tak2007}, using the Large Velocity
Gradient (LVG) approximation for an expanding sphere. The physical
conditions in the molecular gas associated with the filaments and
knots in the Crab Nebula were measured by \cite{Loh2012}. These
authors employed emission lines from ortho-H$_2$ to derive (molecular)
temperatures in the range $\sim 2200-3200$~K. In addition, emission
lines from atomic S$^+$ were used to estimate $n_e\sim
1400-2500$~cm$^{-3}$, from which total hydrogen densities ($n_{\rm
  H}=n({\rm H^+})+n({\rm H})+2n(\rm{H_2}$)) in the range $\sim
14,000-25,000$ were inferred \citep{Loh2012}. As electron-impact rate
coefficients for dipolar transitions are typically $10^4-10^5$ larger
than those of neutrals, collisions with H and H$_2$ can be safely
neglected. Assuming that the proton (H$^+$) fraction is small in the
filaments, electrons should be the dominant colliding species, with a
fraction $n_e/n_{\rm H}\sim 0.1$. The electron density and temperature
were fixed in our calculations at $n_e=2000$~cm$^{-3}$ and
$T=3000$~K. The line width (FWHM) is unresolved with Herschel/SPIRE
but it was taken here as 20~kms$^{-1}$, as expected in the molecular
gas of the nebula \citep{Richardson2013}. Finally, the column density
of $^{36}$ArH$^+$ was varied from $10^{11}$ to $10^{13}$~cm$^{-2}$.

\begin{figure}
\begin{center}
	\rotatebox{-90}{\includegraphics*[scale=0.35]{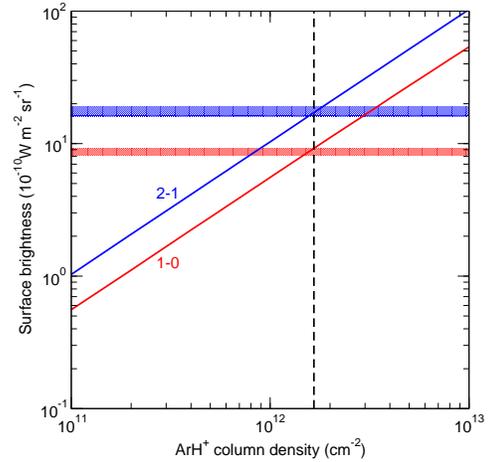}}
\end{center}
\caption{Surface brightness (in 10$^{-10}$Wm$^{-2}$sr$^{-1}$) for the
  $j=1-0$ and $2-1$ rotational lines of $^{36}$ArH$^+$ from the Crab
  Nebula as function of column density. The blue and red hatched zone
  show the observational results of Barlow et al. (2013) at the
  position of the SLW C4 and SSW B3 SPIRE detectors.}
\label{fig:bright.eps}
\end{figure}

Our results are presented in Fig.~\ref{fig:bright.eps} for a position
in the Crab Nebula where both the $j=1-0$ and $2-1$ lines were
detected (SLW C4 and SSW B3 detectors, respectively, see
\cite{Barlow2013}). It is found that the observed surface
brightnesses of the two rotational lines can be reproduced
simultaneously by our model for a column density of $\sim 1.7\times
10^{12}$~cm$^{-2}$. This value is in good agreement with the 
  lower limit of \cite{Barlow2013}, as expected from our higher
  electron density and larger $0\rightarrow 1$ collisional rates. It
should be noted that the surface brightesses at other positions in the
nebula are similar, within a factor of $\sim 3$, suggesting similar
column densities. We can also notice that the surface brightnesses
increase linearly with the column density, as expected in the
optically thin regime (opacities are much lower than unity).

\begin{figure}
\begin{center}
	\rotatebox{-90}{\includegraphics*[scale=0.35]{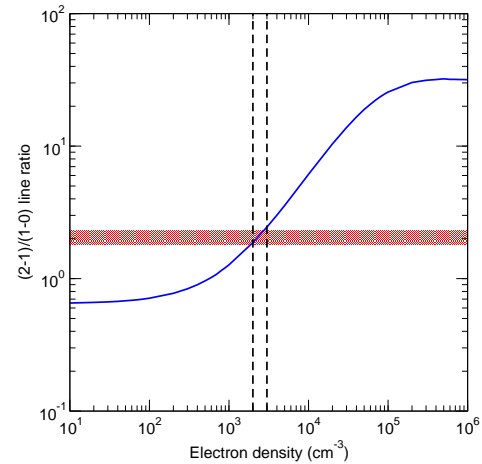}}
\end{center}
	\caption{Plot of the $(2-1)/(1-0)$ line surface brightness
          ratio predicted for a $^{36}$ArH$^+$ column density of
          $2\times 10^{12}$~cm$^{-2}$. The red hatched zone shows the
          observational result of Barlow et al. (2013) at the position
          of the SLW C4 and SSW B3 SPIRE detectors.}
	\label{fig:ratio.eps}
\end{figure}

In order to test the influence of the electron density, we show in
Fig.\ref{fig:ratio.eps} the sensitivity of the $(2-1)/(1-0)$ line
emission ratio. The ratio derived by \cite{Barlow2013} at the above
position in the nebula is $\sim 2$ and it was found to be very similar
($\sim 2.5$) on a bright knot. In our LVG calculations the temperature
is still fixed at $T=3000$~K and the column density at
$N(\rm{ArH}^+)=2\times 10^{12}$~cm$^{-2}$. On Fig.~\ref{fig:ratio.eps}
it is observed that at high electron density ($n_e>10^5$~cm$^{-3}$),
when the rotational populations are thermalized (i.e. at LTE), the
line ratio is $\sim 31$. The ratios observed in the Crab Nebula
therefore suggest a strongly non-LTE excitation of ArH$^+$. We also
observe that the line ratio strongly decreases with decreasing
electron density. Our model thus requires electron densities in the
range $2-3\times 10^3$~cm$^{-3}$ to reproduce the observed ratio, in
very good agreement with the values derived by \cite{Loh2012} from
S$^+$ lines. The line ratio was also found to depend only very weakly
on column density (in the range $10^{11}$ to $10^{13}$~cm$^{-2}$) and
temperature (in the range 2000-3000~K). As a result, the $(2-1)/(1-0)$
emission line ratio of ArH$^+$ is found to provide a robust and
accurate measurement of the electron density in hot and highly ionized
region of the molecular ISM.

\section{Conclusions}

In this paper, we present electron-impact rotational rate
coefficients for diatomic hydride cations of great astrophysical
interest. A complete set of data has been placed in the BASECOL
\citep{jt547} and LAMDA \citep{Schoier2005} databases. These data
comprise electron-impact rotational excitation and de-excitation rate coefficients for
HeH$^+$, $^{12}$CH$^+$, $^{13}$CH$^+$, $^{36}$ArH$^+$, $^{38}$ArH$^+$ and $^{40}$ArH$^+$, for all
transitions $\Delta J = 1 - 8$ involving initial states of $J = 0 -
10$ for excitation and $J = 11 -1$ for de-excitation. These (de)excitation rates should provide useful inputs for
models of a variety of different astronomical environments were
molecular ions are formed. We have in particular employed the
$^{36}$ArH$^+$ data to model the 1-0 and 2-1 emission lines observed
by the Herschel Space Observatory towards the Crab Nebula. These data
have allowed us to derive both the $^{36}$ArH$^+$ column density and
the electron density in this source.
Finally we note that both $^{36}$ArH$^+$ and $^{38}$ArH$^+$
have recently been detected in an extragallictic source \citep{muller}, 
the collisional data provided
here should prove useful in interpretating such observations.

\section*{Acknowledgements}

This work has been supported by an STFC CASE studentship, grant number
ST/K004069, for JRH, the Agence Nationale de la Recherche
(ANR-HYDRIDES), contract ANR-12-BS05-0011-01 and by the CNRS national
program "Physico-Chimie du Milieu Interstellaire". Philippe Salom\'e
is acknowledged for useful discussions on the Crab Nebula.

%\begin{section}*{References}
%\References
\bibliographystyle{mn2e}
%\bibliography{journals_phys,n2,jtj,R-Matrix,Hydrides}
%\end{section}

\end{document}